\documentclass[sigconf, nonacm, screen]{acmart}

\usepackage{bm}
\usepackage{color}
\usepackage{subcaption}
\usepackage{multirow}
\usepackage{float}
\usepackage{algorithm}
\usepackage{algorithmic}
\usepackage{cleveref}
\usepackage{xcolor}
\usepackage[colorinlistoftodos]{todonotes}
\usepackage{tabularx} 
\usepackage{booktabs} 
\usepackage{chngcntr} 

\definecolor{lightlightgray}{gray}{0.85}
\definecolor{lllgray}{gray}{0.9}
\definecolor{llllgray}{gray}{0.95}
\crefname{equation}{Eq.}{Eq.}
\crefname{section}{Section}{Sections}
\crefname{subsection}{Section}{Sections}
\crefname{subsubsection}{Section}{Sections}
\crefname{figure}{Fig.}{Figs.}
\crefname{table}{Table}{Tables}
\crefname{subfigure}{Fig.}{Figs.}
\crefname{algocf}{Algorithm}{Algorithms}

\newtheorem*{hypothesis}{Hypothesis}

\usepackage{enumitem}
\setlist{after=\vspace{0\baselineskip},leftmargin=12pt}
\copyrightyear{2024}
\acmYear{2024}
\setcopyright{rightsretained}
\acmConference[WWW '24]{Proceedings of the ACM Web Conference 2024}{May 13--17, 2024}{Singapore, Singapore}
\acmBooktitle{Proceedings of the ACM Web Conference 2024 (WWW '24), May 13--17, 2024, Singapore, Singapore}\acmDOI{10.1145/3589334.3645634}
\acmISBN{979-8-4007-0171-9/24/05}

\begin{document}

\title{Exit Ripple Effects: Understanding the Disruption of Socialization Networks Following Employee Departures}

\author{David Gamba}
\affiliation{
  \institution{University of Michigan}
  \country{Ann Arbor, MI, USA}
  }
\email{gamba@umich.edu}

\author{Yulin Yu}
\affiliation{
  \institution{University of Michigan}
  \country{Ann Arbor, MI, USA}
  }
\email{yulinyu@umich.edu}

\author{Yuan Yuan}
\affiliation{
  \institution{Purdue University}
  \country{West Lafayette, IN, USA}
  }
\email{yuanyuan@purdue.edu}

\author{Grant Schoenebeck}
\affiliation{
  \institution{University of Michigan}
  \country{Ann Arbor, MI, USA}
  }
\email{schoeneb@umich.edu}

\author{Daniel M. Romero}
\affiliation{
  \institution{University of Michigan}
  \country{Ann Arbor, MI, USA}
  }
\email{drom@umich.edu}

\renewcommand{\shortauthors}{David Gamba, Yulin Yu, Yuan Yuan, Grant Schoenebeck, \& Daniel Romero}

\begin{abstract}
Amidst growing uncertainty and frequent restructurings, the impacts of employee exits are becoming one of the central concerns for organizations. Using rich communication data from a large holding company, we examine the effects of employee departures on socialization networks among the remaining coworkers. Specifically, we investigate how network metrics change among people who historically interacted with departing employees. We find evidence of ``breakdown" in communication among the remaining coworkers, who tend to become less connected with fewer interactions after their coworkers' departure. This effect appears to be moderated by both external factors, such as periods of high organizational stress, and internal factors, such as the characteristics of the departing employee. At the external level, periods of high stress correspond to greater communication breakdown; at the internal level, however, we find patterns suggesting individuals may end up better positioned in their networks after a network neighbor's departure. Overall, our study provides critical insights into managing workforce changes and preserving communication dynamics in the face of employee exits.
\end{abstract}




\maketitle

\section{Introduction}

Organizations are dynamic entities where personnel changes are an inherent feature. As employees depart, whether through voluntary resignation or enforced layoffs, there are consequences on the structural and functional aspects of the organization. Prior literature indicates that these departures may impact employee morale, knowledge transfer, productivity, and other organizational outcomes \cite{krackhardtWhenFriendsLeave1985, abbasiTurnoverRealBottom2000,sheehanEffectsTurnoverProductivity1993, ongoriReviewLiteratureEmployee2007}. 

Quantifying the influence of an individual's departure on an organization has been challenging, often prompting researchers to depend on qualitative assessments \cite{klotzSayingGoodbyeNature2016}, narrowing to key employees \cite{charfeddineInfluenceCEODeparture2015} or to relating departure turnover rates to macro impacts on group and company performance \cite{abbasiTurnoverRealBottom2000,sheehanEffectsTurnoverProductivity1993, ongoriReviewLiteratureEmployee2007}. In this study, we focus on the impact on socialization networks, specifically the network interactions among remaining employees. For example, consider a scenario where Alice, Bob, and Charlie interact regularly at work. If Alice exits the company, how does this change the communication dynamics between Bob and Charlie? While our approach does not directly measure tangible outcomes like productivity or revenue, which are difficult to attribute to a single departure, it provides a clear and quantifiable way to understand the implications of personnel changes on interaction patterns and networks. Such interaction networks have proved indispensable for multiple contexts within companies, such as development of organizational advantage \cite{argoteKnowledgeTransferBasis2000, nahapietSocialCapitalIntellectual1998}, collaborative task development \cite{krepsCorporateCultureEconomic1990,elliottCorporateCultureOrganizational2023, charpignon2023navigating}, and for the well-being of employees \cite{podolnyResourcesRelationshipsSocial1997}.

Furthermore, the uniqueness of our dataset allows us to scrutinize how departures influence socialization networks during periods of high stress. Notably, our data spans two distinct periods: one where the firm experienced stress and ambiguity, and another where it operated under more typical conditions. Drawing on findings from previous research which outlines how intra-organizational networks change with stress and ambiguity \cite{romeroSocialNetworksStress2016, driskellStressPerformanceDecision2014}, we analyze the interaction effects between the external stress level of the firm and individual departures. We find systematic differences: compared to low-stress periods, departures during high-stress periods  are associated with less communicative groups in the company but are also associated with structural patterns that benefit remaining employees.

In addition to external factors, we investigate the heterogeneity of the departing employee's attributes. Gender, for instance, can influence communication styles and collaborative tendencies \cite{forretNetworkingBehaviorsCareer2004, chowDoesGenderManager2011}, while some communication attributes, such as network closure within ones social network, correlate with organizational knowledge transfer and influence \cite{reagansNetworkStructureKnowledge2003, davenportStrategiesPreventingKnowledgeloss2006}. As such, we are particularly interested in exploring how these factors affect socialization dynamics post-departure.

Our research questions are thus: \textbf{RQ1} What is the effect of an employee's departure on the socialization networks of their prior contacts? \textbf{RQ2} How is this effect different during organizational high-stress periods? and \textbf{RQ3} How are the attributes of the departing employee, such as their volume of communication or seniority, related to the response in communication dynamics of their prior contacts?

To frame our analysis, we draw on past research on social capital operationalized through networks \cite{podolnyResourcesRelationshipsSocial1997, burtStructuralHolesSocial1992}. We employ a large-scale dataset of internal communications from a major company among \textasciitilde{}100K employees. From these data, we track changes in communication dynamics of departing employees' contacts on a weekly basis, measuring various attributes associated with the group and individual dynamics over a period before and after departures take place. Following a model-based matched comparison approach, we assess the relationship between departures and changes in the socialization dynamics relative to a control group.

Our main contributions are as follows: \textbf{(1)} 
To the best of our knowledge, this is the first large scale analysis of the dynamics of networks after node removal in an organizational context that also considers the effect of high stress organizational environments.\textbf{(2)} Our results suggest a breakdown of socialization among the remaining members after a departure. The breakdown is marked by less connectivity, volume, cohesiveness, and efficiency when looking at communications only between the employees who used to interact with the departing employee  (group perspective) and their interactions within the larger organization (individual perspective). \textbf{(3)} We compare the response to departures between periods of high and low stress. Notably, departures during high-stress periods are associated with more detrimental structural consequences to groups but beneficial for individual employees. \textbf{(4)} Our research provides valuable insights into the field of organizational research by underscoring the ripple effects of a departure.

\section{Theoretical Framework and Related Work}
Here, we build a conceptual framework for our research informed by literature on organizational departures, socialization and social capital, and node removal in complex networks.

\paragraph{Understanding Socialization Through Social Capital and Network Structures}

\begin{figure}
  \centering
  \includegraphics[width=0.47\textwidth]{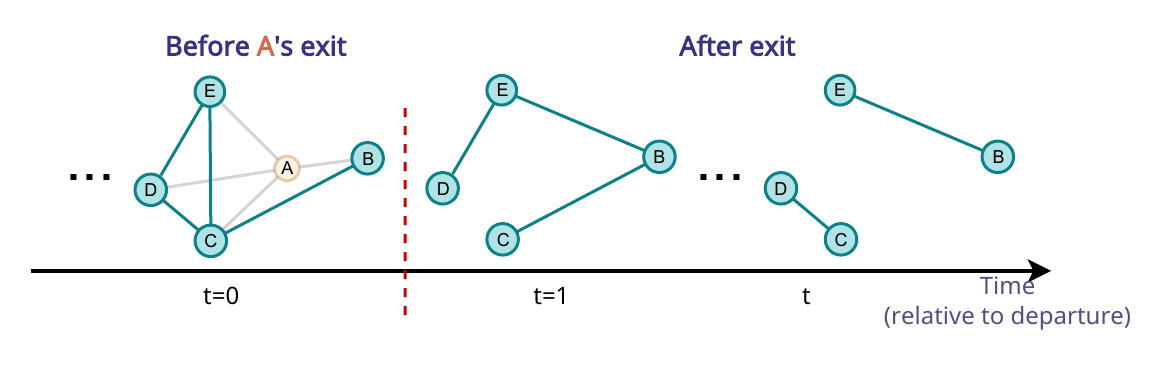}
  \Description[Example of group tracking]{Example of group tracking}
  \caption{We track the evolution of the interactions of neighbors of a departing employee. Example shows how group interactions change after A's departure and ends up disconnected into silos. Here we only show the group perspective for simplicity.} 
  \label{fig:tracking}
\end{figure}

In our analytical framework, we use interactions in the communication data as a proxy for employee socialization. Socialization is crucial to onboarding and it is instrumental for employee effectiveness \cite{bauerOrganizationalSocializationEffective2011a, fangOrganizationalSocializationProcess2011, feldmanMultipleSocializationOrganization1981}. We argue that when people socialize within companies, they're building and using social capital. One common lens to operationalize social capital is through networks \cite{podolnyResourcesRelationshipsSocial1997,burtStructuralHolesSocial1992}. This is the lens we adopt in the present study. From the data, we construct communication networks. In these networks, we focus on the interactions within what we define as the \textbf{socialization set} of departing employees. A person in this set is an organization member whom the departing employee would interact with in normal day-to-day work circumstances. The interaction might be regular (as teammates or supervisors) or could be sporadic. The inclusion of these sporadic ties is informed by the concept of "weak ties" \cite{granovetterStrengthWeakTies1973}, which are instrumental in information flow in organizations \cite{burtCooperationNetwork2022, gargiuloTrappedYourOwn2000}

\paragraph{Dual perspectives on socialization and interactions}

Current literature does not clearly connect organizational departures with employee socialization nor the mediation effect of a network node's removal. Nevertheless, it hints at reactions at different levels: 
At the broader company level, negative effects include decreases in productivity and a hit on the morale of remaining employees after key employee departures \cite{abbasiTurnoverRealBottom2000,sheehanEffectsTurnoverProductivity1993, ongoriReviewLiteratureEmployee2007}. Some research finds positive effects in performance when figures such as top managers leave \cite{hausknechtCollectiveTurnoverGroup2011, denisPerformanceChangesFollowing1995}. We contrast these \emph{group} effects with other parts of the literature that focus on \emph{individuals}. Looking at single employees, departures can induce increased commitment and satisfaction from the remaining collaborators \cite{krackhardtWhenFriendsLeave1985}, or lead to positive career outcomes such as increased job mobility  \cite{pogeFillingGapConsequences2022, andersonLeftUnderstandingCareer2023a}.

The literature suggests departures might affect groups and individual employees differently. We account for this by adopting two central perspectives of interaction given the socialization set. The \textbf{group perspective} sees the socialization set of a departing employee ($e$) as a cohesive unit, focusing on the communication solely among its members. These interactions might be crucial even when they cross team boundaries as they are valuable for collaboration and socialization \cite {clarkegarciaTiesLeadersTeams2014a, feldmanMultipleSocializationOrganization1981}. Complementing this, the \textbf{individual perspective} sees all interactions of members of the socialization set, including those beyond the socialization set's boundaries, focusing on their entire network within the organization.

\paragraph{Network measures of social capital}

We leverage two key bodies of literature to establish measures for the networks we examine and to align our findings with existing research. The first is the concept of social capital from a network perspective, which investigates how network structures influence organizational outcomes. The second is the research on node removal within networks. 

Two key concepts in social capital via networks are network closure and structural holes. Closure has been positively associated with team performance \cite{balkundiTiesLeadersTime2006a,clarkegarciaTiesLeadersTeams2014a} and with individual sense of cohesion and well-defined expectations \cite{podolnyResourcesRelationshipsSocial1997, gargiuloTrappedYourOwn2000}. However, it can also instigate a sense of coercion and self-segregation \cite{burtCooperationNetwork2022, podolnyResourcesRelationshipsSocial1997}. Structural holes create disconnected network spaces, offering bridging opportunities. Employees acting as bridges to fill network gaps often achieve better career outcomes \cite{gargiuloTrappedYourOwn2000, burtStructuralHolesSocial1992} and allow for accessing disparate sources of information \cite{reagansNetworkStructureKnowledge2003, burtStructuralHolesSocial1992}. Additionally, employees bridging structural holes display increased flexibility to organizational position changes \cite{gargiuloTrappedYourOwn2000}. 

We also draw from literature on node removal on complex networks. Here characteristics that are important to understand in a group are efficiency, referring to information flow, redundancy \cite{latoraEfficientBehaviorSmallworld2001, boccalettiComplexNetworksStructure2006b}, and the size of the largest component \cite{bellingeriLinkNodeRemoval2020,iyerAttackRobustnessCentrality2013, bellingeriEfficiencyAttackStrategies2014, wandeltComparativeAnalysisApproaches2018, erdosEvolutionRandomGraphs2011}. These properties are often considered to evaluate the impact of node removal on a network. We adopt these measures in our analysis. 

\paragraph{Shocks to networks, turbulent environments}
Our review also extends to the literature detailing shocks and high-stress environments for networks. We use these insights to form hypotheses regarding the effect of employee departures during times of high stress. 

Researchers have identified a phenomenon referred to as `turtling-up' when networks are subjected to abrupt shocks. The network restricts ties with out-group members, instead intensifying the cohesion of within-group ties \cite{romeroSocialNetworksStress2016, romero2019social}. In the face of changing environments, for instance, the shift to remote work, networks tend to become more compartmentalized \cite{yangEffectsRemoteWork2022a}. When observing the evolution of within-group social dynamics during periods of heightened uncertainty, notable changes in structure and tie composition among employees become apparent: Relationships are more prone to experience conflict, cohesion of group declines, and hierarchical structures among group members emerge \cite{srivastavaIntraorganizationalNetworkDynamics2015a, giannoccaroTeamResilienceComplex2018, driskellTeamsExtremeEnvironments2018}. Researchers have broken down ties into formal, semi-formal, and informal networks and found that structural shifts often involve a decline in formal network ties, counterbalanced by an increase in semi-formal and informal ties \cite{srivastavaIntraorganizationalNetworkDynamics2015a, driskellTeamsExtremeEnvironments2018}.

\subsection{Hypotheses} \label{sec:hypotheses}

\paragraph{Effect of an employee departure on the interactions within their socialization set}
From a group perspective, morale and efficiency could drop significantly after a departure \cite{abbasiTurnoverRealBottom2000,sheehanEffectsTurnoverProductivity1993, ongoriReviewLiteratureEmployee2007}, which suggests decline in communication and efficiency \cite{latoraEfficientBehaviorSmallworld2001, boccalettiComplexNetworksStructure2006b}. Furthermore, the departing employee might mediate the stability of a triad between coworkers. This connection might be compromised due to a node's removal \cite{granovetterStrengthWeakTies1973}, creating the potential for fragmentation of the group into subgroups. From an individual perspective, some individuals might show increased commitment, fostering unity within their immediate circles \cite{krackhardtWhenFriendsLeave1985}. Alternatively, others might display individual advantage patterns \cite{andersonLeftUnderstandingCareer2023a, pogeFillingGapConsequences2022}, such as an increase in connections, a reduction in their clustering, and an increase in brokerage positions within the company \cite{burtStructuralHolesSocial1992, reagansNetworkStructureKnowledge2003}.

\begin{hypothesis}[H1.1]\label{hyp:1.1}
From a group perspective, the interactions of the socialization will break apart, marked by a decrease in communications and the number of connections within the group, as well as by a decrease in cohesiveness and closeness of the group. 
\end{hypothesis}

\begin{hypothesis}[H1.2a]\label{hyp:1.2a}
From an individual perspective, on average, members of the socialization set will display an increased commitment effect reflected in increased individual clustering and fewer structural holes in their ego networks.
\end{hypothesis}

\begin{hypothesis}[H1.2b]\label{hyp:1.2b}
Alternatively, from an individual perspective, on average, members of the socialization set will display an individual advantage effect, where they increase their connections and end up in positions with increased structural advantage, which leads to lower clustering and becoming bridges of structural holes \cite{andersonLeftUnderstandingCareer2023a, pogeFillingGapConsequences2022}.
\end{hypothesis}

\paragraph{Departures under high-stress environment}

According to literature, during high-stress periods, groups tend to isolate, forming disconnected components\cite{yangEffectsRemoteWork2022a}. This implies a departure could worsen communication disruption from the group perspective, exacerbated by the stress-induced tendency towards disconnection.

\begin{hypothesis}[H2.1]\label{hyp:2.1}
 During a high-stress period, at the group level, we will observe increased group breakdown, evidenced by larger effect sizes in the measures mentioned above.
\end{hypothesis}

The literature about individual employee adaptation in uncertain periods suggests individuals preserve and strengthen informal ties \cite{srivastavaIntraorganizationalNetworkDynamics2015a, driskellTeamsExtremeEnvironments2018}. Thus, following a departure, we anticipate that the resulting uncertainty escalated by both the environment and the employee's exit prompts individuals to maintain or seek diverse connections. This dynamic could also diminish the clustering of connections, enrich diversity, and foster increased network brokerage. However, alternatively, the effect of stress and hit to morale seen in turbulent times \cite{mujtabaLayoffsDownsizingImplications2020} might be significant enough that the individuals turtle-up showing increased clustering and less communication \cite{romeroSocialNetworksStress2016}. This leads us to propose two competing hypotheses.

\begin{hypothesis}[H2.2a]\label{hyp:2.2a}
 During a high-stress period, at the individual level, we will observe increased brokerage patterns. 
\end{hypothesis}

\begin{hypothesis}[H2.2b]\label{hyp:2.2b}
 During a high-stress period, at the individual level, we will observe increased isolation patterns.
\end{hypothesis}

\section{Data and Methods} \label{sec:methods}
We now proceed to describe the data sources and methodology in order to test the hypotheses established in the previous section.

\subsection{Data Context}\label{sec:methods:data}
Our research primarily investigates socialization patterns within a large company by using its internal communications data. These data cover periods of high and low stress. The high-tress periods follow significant regulatory changes that occurred early in 2021. The new regulations could potentially make a significant part of the company's operations illegal. The threat and later imposition of these regulations created a sense of uncertainty within the company, resulting in significant workforce upheaval and attrition. Such effects may have influenced the company's internal communication practices during this period.

A vast portion of our data comes from the company's dominant communication network, which is an instant messaging tool akin to Slack. These data span the year 2021 and contain ~5M weekly interactions among approximately 120,000 employees. To ensure that our data only covers regular workday interactions, we excluded periods like holidays, which could predictably influence communication patterns and introduce bias.

\subsection{Networks} \label{sec:methods:networks}

\paragraph{Constructing the Weekly Communication Network}
Similar to previous studies \cite{dechoudhuryInferringRelevantSocial2010,yuLargeScaleAnalysisNew2023, yangEffectsRemoteWork2022a}, we build weekly graphs, denoted as $\mathcal{G}^w = (\mathcal{V}^w, \mathcal{E}^w, \mathcal{W}^w)$, where $w$ denotes the week. This graph has nodes representing the set of all employees, $\mathcal{V}_w$, that have communicated within that week. Correspondingly, the edges in $\mathcal{G}^w$, denoted as $\mathcal{E}^w$ represent interactions between employees and have weights $w_{ij} \in \mathcal{W}$. Every edge weight $w_{ij}$ is constructed with the aggregate volume of communications between $a$ and $b$ during that week. In our networks, we include both pairwise interactions (direct messages) as well as group interactions. The latter accounts for how groups maintain socialization and interaction in non-dyadic channels \cite{yangEffectsRemoteWork2022a}. Specifics on weight calculations can be found in the \cref{app:networks_construction}.

\paragraph{Constructing Socialization Sets}
We denote the socialization set of an employee $e \in \mathcal{V}$ as $SS_e \subset \mathcal{V} = \bigcup\nolimits_{w=t_e^*-10}^{t_e^*-6} \Gamma_{\mathcal{G}^w}(e)$, with $t^*_e$ marking the calendar week of departure of $e$. We include only members that interact during weeks $[t^*-10, t^*-6]$ for two reasons. First, we aimed to exclude any interactions that could form part of an offboarding process from the socialization set. As we lack specific data about whether departures were voluntary or involuntary, we decided to adopt a buffer period of 6 weeks prior to the departure, given that the company suggests that leave notices are given one to two months prior to voluntary departures. Then, we select a 'freeze' window of 4 weeks from the buffer $[t-6 -4, t-6]$ to include people that have interacted during this window as the socialization set. We have also completed a robustness test where vary the freeze size. We find the results of the main analysis are qualitatively similar. 

\paragraph{Estimating Employee Departures}

We identify an employee's departure when they cease to participate in IM communications, leveraging the prevalence of the IM channel as the defacto communication channel in the company. This is a reliable measure until 2021 since we IM data until mid-2022. Thus, eliminating possible false positives due to vacations or similar phenomena. We identified ~40K employee departures in 2021.

\paragraph{Representing group and individual perspectives}
We define: 

\begin{description}
    \item[Group perspective] as $G^{e,w}_{grp} = \mathcal{G}^w[SS_e]$, which is the induced graph restricted to the socialization set on a given week. This represents the interactions only between members of a departing employee's socialization group.
    \item [Individual perspective] as $\{G^{e',w}_{ind}\}_{e'\in SS_e}$, where \\ $G^{e',w}_{ind} = \mathcal{G}^w[{e'\cup\Gamma_{\mathcal{G}^w}(e')}]$ is the ego network of $e'$, where $e'$ is a member of departing employee $e$'s socialization set. Thus, the individual perspective considers the ego network of each member of $e$'s socialization set, including communication with any employee in the organization, beyond other members of $e$'s socialization set. 
    
\end{description}

\subsection{Measures of socialization} \label{sec:methods:measures}

\begin{table}
\centering
\caption{Grouping of measures and constructs}
\label{tab:measures}
\begin{tabularx}{\columnwidth}{>{\hsize=.45\hsize}X>{\hsize=.6\hsize}X}
\toprule
\textbf{Construct Measured}         & \textbf{Metric(s)}                     \\ \midrule
\textbf{Group Perspective} &                  \\
\hspace{1em}Interaction Intensity   & Connections, Volume        \\
\hspace{1em}Cohesion                & Closure, Components \\
\hspace{1em}Efficiency              & Largest component share, Closeness             \\ \midrule
\textbf{Individual Perspective} &              \\
\hspace{1em}Interaction Intensity          &  Connections (ind.), Volume (ind.)          \\
\hspace{1em}Structural Advantage    & Diversity                              \\
\hspace{1em}Entrenchment & Clustering (ind.)  
\\ \bottomrule
\end{tabularx}%
\end{table}

We extract relevant metrics from the aforementioned individual and group perspectives on a weekly basis for up to 32 weeks centered around an employee's departure. This process provides us with time-series data, $f^{e,m}(t)$, where $m$ represents the calculated metric, $e$ the socialization set index tied to the departing employee, and $t$ the time relative to the employee's departure. Following, we define the metrics. \cref{tab:measures} contains a grouping of metrics based on the constructs we aim to measure.

For the \textbf{group perspective}, we calculate the following:

\begin{description}
    \item[Closeness] Measures how well connected the group is and it is akin to a measure of efficiency in the group \cite{latoraEfficientBehaviorSmallworld2001, boccalettiComplexNetworksStructure2006b}. It is calculated as the average inverse distance of all pairs of nodes in  $G^{e,w}_{grp}(G,e)$.
    \item[Closure] Measures group cohesiveness via triadic closure \cite{granovetterStrengthWeakTies1973}; measured by the average clustering coefficient of $G^{e,w}_{grp}$.
    \item[Components] Measures disconnected silos of interaction within the group. Calculated as the number of components in $G^{e,w}_{grp}$.
    \item[Largest component share] Measures network robustness \cite{iyerAttackRobustnessCentrality2013, wandeltComparativeAnalysisApproaches2018, bellingeriEfficiencyAttackStrategies2014,erdosEvolutionRandomGraphs2011}. Calculated as $|G^{e,w}_{grp, 0}|/|G^{e,w}_{grp}|$ where $G^{e,w}_{grp, 0}$ denotes the largest component of $G^{e,w}_{grp}$.
    \item[Connections] Measures the volume of pairwise interactions among the members of the socialization set. Defined as the number of edges in $G^{e,w}_{grp}$ normalized by $|G^{e,w}_{grp}|$.
    \item[Volume] Measures aggregate volume of interactions. Calculated as the sum of weighed edges in $G_{grp}(G,e)$ normalized by $|G^{e,w}_{grp}|$.
    \item[N active] The number of active socialization set members. It allows us to assess if other members of the socialization also leave or become inactive. Defined as $|G^{e,w}_{grp}|$
\end{description}

For the \textbf{individual perspective}, we calculate metrics for each $G^{e',w}_{ind} \in \{G^{e',w}_{ind}\}_{e'\in SS_e}$, which we then average to get one aggregate estimate. We describe now each measure in terms of each $G^{e',w}_{ind}$:

\begin{description}
    \item[Clustering] Represents the embeddedness of $e'$ in their own network \cite{colemanSocialCapitalCreation1988, podolnyResourcesRelationshipsSocial1997, gargiuloTrappedYourOwn2000}. Calculated as the local clustering of the node $e'\in SS_e$ in the network $ego(e’, G)$.  
    \item[Connections (ind.)] Represents how connected the employee is. Calculated for a given $e' \in SS_e$ as $|G^{e',w}_{ind}|$.
    \item[Volume (ind.)] Represents overall employee communication. We take the weighted edge sum of $e'$ in $G^{e',w}_{ind}$.
    \item [Diversity] Is an indication of how many structural holes $e'$ bridges  \cite{dechoudhuryInferringRelevantSocial2010, uganderStructuralDiversitySocial2012}. It is calculated as the number of components in the graph $G^{e',w}_{ind} / e'$, ie. the components of the ego network of $e'$ removing $e'$ and its edges.
\end{description}

\subsection{Matching}
The inherent nature of corporate dynamics, coupled with prevailing uncertainties, can result in diverse and dynamic socialization patterns within a company. It is paramount to account for these externalities when analyzing the effect of an employee's departure on their socialization set. For this reason, we incorporate a matching design to generate a control group that serves as a comparison for socialization sets of non-departing employees. For each socialization set, we find a set of $k$ matches that are similar in departing employee network attributes and socialization set metrics (See \cref{app:matching_details} for a list of the matching attributes). We generate an estimate of these attributes given an employee $e$ by averaging the metrics over a defined period of time $[t^*_e-10, t^*_e-6]$, which corresponds to the same period of time on which we defined the socialization set $SS_e$ relative to $e$'s departure. We use k-nearest neighbors (kNN) as the matching algorithm. Further details about this procedure are in the \cref{app:matching_details}.

\subsection{Models} \label{sec:methods:model}

\paragraph{Model definition}
We develop a model-based approach to quantify the change in the metrics of the socialization sets after the departure of their corresponding ego employees. For each metric denoted as $f^{m}$, we fit a model to encapsulate its dynamics; it models the time progression of the metrics with linear trends relative to the departure timing of the socialization set’s ego $e$ \cite{fitzmauriceAppliedLongitudinalAnalysis2011}. Our model
incorporates a linear splines basis, which allows for modeling changes in value and trend of the metrics after departure \cite{seberNonlinearRegressionModels2015}, random effects to account for the variability between different socialization sets  \cite{raudenbushHierarchicalLinearModels2002}, and uses the matched socialization groups to perform before-after contrast comparison of response in metrics using a control group \cite{ryanWhyWeShould2015}. The model is expressed as follows:

\begin{equation} \label{eq:model-main}
f^m_{e,t} \sim A_e \times (t + hinge(t) + jump(t))
+ controls_e + \eta_{e,t},
\end{equation}

where we have functions that define the time basis: $hinge(t) = t * \vmathbb{1}(t>0)$ denotes a change in slope and $jump(t) = \vmathbb{1}(t>0)$ denotes a discontinuity in value at time $t$. Here, $e$ denotes each socialization set $SS_e$, which is itself defined by the departing ego $e$, $t$ corresponds to the relative time to departure, $A_e$ is an indicator with a value of $1$ if the ego employee of the socialization set $e$ is a departing employee (treatment group) and $0$ is they are in the matched control group. $\eta_{e,t}$ encodes random effects by departing ego/socialization set, which accounts for baseline metric variations among distinct employee departure groups. Finally, using the interaction between the treatment indicator and time basis, we can compare the treatment and control groups' relative discontinuity and slope differences using marginal estimates, which we describe in the next section. For model fitting, we transform target variables to allow for comparison between different metrics and adjust for heavy-tailed variables. Further details can be found in the \cref{app:modeling_details}. 

\paragraph{Model estimates} 
We quantitatively measure the response of socialization sets using estimates extracted from  our fitted model parameters (details of expressions and calculations are in the \cref{app:modeling_details}):

\begin{description}
    \item[Value DiD] Denoted as  $DiD_{val}(\hat{f})$, contrasts how the metric changes from the pre-departure to departure period compared to the control group. We calculate it as the difference (between groups) of the difference in each group comparing before and after values of the estimates of a metric. For example, a value of 0.1 in the metric $f$ means that compared to the control group, the socialization sets of departing employees change 0.1 DM (standard deviations of the metric) more relative to the control group change.

    \item[Slope DiD] Denoted as $DiD_{slp}(\hat{f})$, captures the metric slope difference between pre and post-intervention for each group estimate. For example, a positive value of this estimate indicates that the change in trend for the treatment group was larger when compared to the control. These estimates reveal whether Value DiD estimates increase or decrease over time. 
    
\end{description}

Using the two estimates, we identify both a difference in the value of a metric after departure and a trend, which indicates if this difference changes over time.

\paragraph{Assessing the effect of uncertainty}

\begin{figure}
    \centering
    \includegraphics[width=0.47\textwidth]{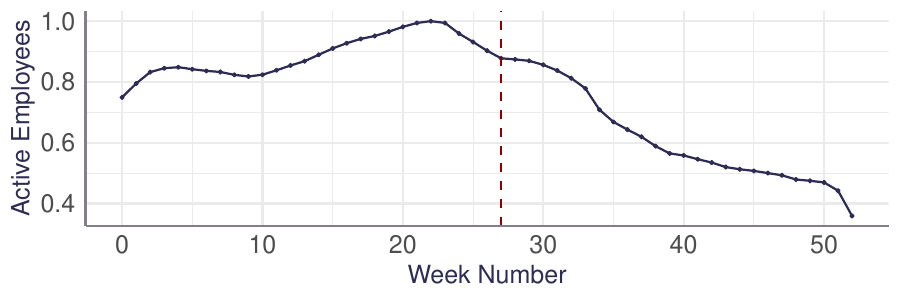}
    \Description[Employee counts for 2021]{Employee counts for 2021}
    \caption{Weekly active employees in the communication data. The red line indicates the week on which the high-stress period starts.  Counts for 2021, normalized by the maximum value.} 
    \label{fig:emp_counts}
\end{figure}

\begin{figure*}
    \centering
    \includegraphics[width=0.975\textwidth]{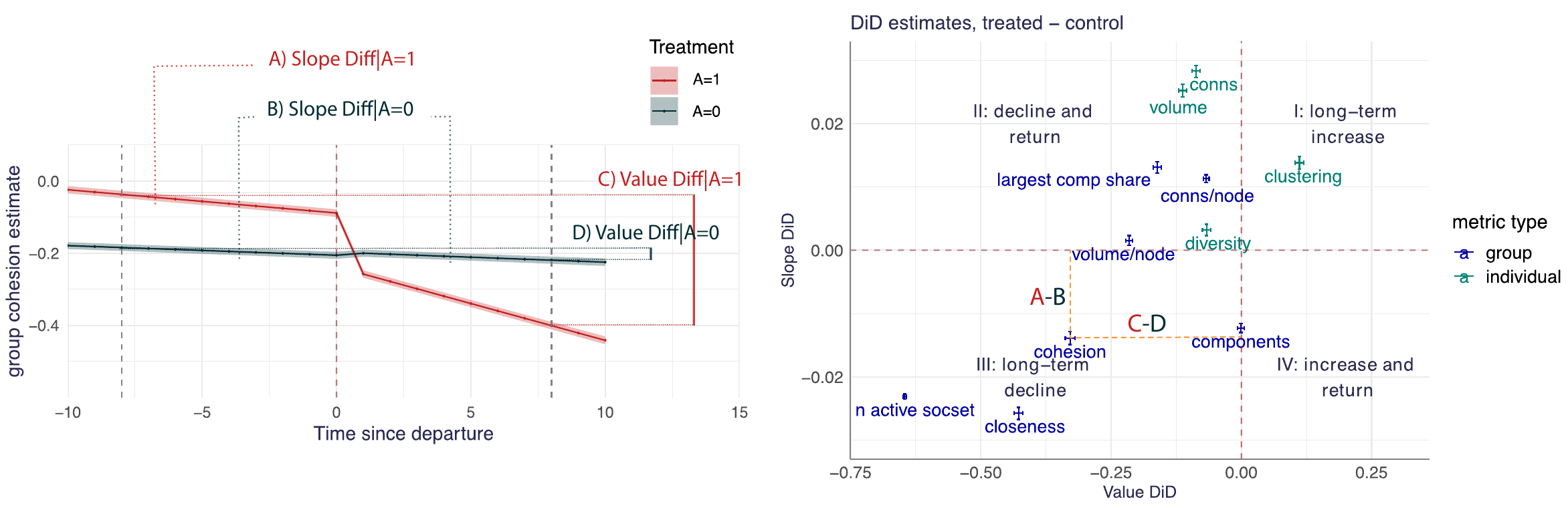}
    \Description[Quadrants for Value and Slope DiD]{Quadrants for Value DiD and Slope DiD}
    \caption{L) Marginal estimates for group cohesion around departure. We also display the value and slope change for each of the treatment groups. The differences in these changes between the groups are the DiDs,  as shown on the right. Note: this is a pictorial example, the detailed estimation procedure to include errors is in \cref{sec:methods:model}. R) Quadrants with model-based DiD estimates of $treated - control$ socialization sets. Most estimates are located in the quadrant of negative value and trend, indicating communication breakdown in both the group and the individual perspectives.}
    \label{fig:main}
\end{figure*}

To study changes in the network structure of socialization sets amidst heightened uncertainty, we exploit the unique aspect of our data, which reflects the ban on company activities implemented by the government. Such policy initiated a period of high uncertainty accompanied by reorganization. We separate the data into two periods, as seen in \cref{fig:emp_counts}. We refer to the period post-implementation of the ban as the high-stress period. We split our data into two sets, with the first week of July as the dividing point. A one-month buffer was applied before and after this date, removing any employee exits within this period from the data. Consequently, we have ~1M observations for ~80K treated and control socialization sets for each period. Then, we apply the same model described in the previous section to each period separately. Using these fitted models, we calculate diff-in-diff estimates for both value and slope, then compare them between the two periods.

\paragraph{Modeling heterogeneous effects}
We also examine differential responses in the socialization set associated with different characteristics of the departing ego individuals. Specifically, we examine ego characteristics of \textbf{leadership status, seniority, and gender}. We also examine communication-related attributes, including \textbf{ego's volume of communication, number of connections, clustering, and structural diversity}. Then, to model the heterogeneous effects, we employ a variation of the previous models where, instead of contrasting treated and control socialization sets, we contrast the effect of the departures across different attribute levels. For example, we contrast how the response to a departure differs when comparing a highly clustered departing employee to a low-clustered employee. Further details of the model are in the \cref{app:modeling_details}.

\section{Results} \label{sec:results}

\subsection{\textbf{RQ1}, Effect of ego’s departure on the socialization set} \label{sec:results:departure}

Addressing RQ1, we calculated the model-based value and slope DiD estimates as outlined in \cref{sec:methods:model}. \cref{fig:main} presents these estimates on two axes: one for value DiD and another for slope DiD. We report these coefficients in terms of the standard deviation of the metric across all observations (DM) and of the rate of change of this unit per week (DM/w) for value DiD and slope DiD, respectively. We refer to the magnitude of the standardized value DiD and slope DiD estimates as effect size. In addition, unless stated otherwise, all values reported in the following sections are significant at a $0.01$ significance level, where we also applied a Bonferroni correction \cite{bonferroniTeoriaStatisticaClassi1936}. Further, we define quadrants that indicate possible situations for the evolution of the metrics contingent on the signs of Value DiD and Slope DiD. We display these quadrants alongside estimates in \cref{fig:main}.

We observe significant coefficients for all the metrics, supporting our hypothesis that socialization sets experience an impact on ego departures when compared to control. We now delve into the nature of this effect from both group and individual perspectives.

From a group perspective, \textbf{we find support for hypotheses H1.1}, we observe a drop in DiD values for metrics associated with interaction intensity (group connections and volume), cohesion (closure) and efficiency(closeness). Closure and closeness were the metrics most impacted within group socialization, with a comparative decline of -0.33 DM and -0.42 DM, respectively. Following the departure of the ego employee, there is also a significant decrease in the number of active members within the socialization set (-0.646 DM). It's important to note that the decrease in interaction intensity is not solely due to the reduction in active members within the socialization set, as these variables were normalized by node count, and the model controls for this factor. This indicates a compounded decrease in group interactions, even with a reduced number of active nodes.

We also examined the slope DiDs to understand temporal dynamics. For the group perspective, connections (0.0113 DM/w) and share of the largest component (0.0131 DM/w) display a positive slope, in contrast to their negative value drops. This suggests the potential convergence of these metric differences over time between the treatment and control groups. Contrasting this, metrics such as group closeness (-0.0257 DM/w), cohesion (-0.0139 DM/w), and number of components (-0.0123 DM/w) show a negative slope DiD, indicating an increasing difference between the treatment and control groups over time. Overall, this implies that treated socialization sets might continue to experience decreasing cohesion and communication volumes.

From the individual perspective, we note a decrease in metrics for individual interactions (connections: -0.0871 DM, volume: -0.1127 DM), and diversity (-0.0668 DM). Compared to other metrics, however, the decrease in diversity is relatively minor. Additionally, there is an uptick in the individual clustering (0.1113 DM). 

Looking at the slope DiD for changes over time, we find that individual employees seem to rebound to the status quo relatively quickly in terms of interaction intensity, as evidenced by the volume and connections metrics displaying the higher value of slope DiD. However, increased clustering within their communication networks (Quadrant I) and stagnant diversity (minor 0.0032 DM/w estimate) become apparent. These observations may imply potential challenges for individuals as brokers of information. Taken together, the tendency towards increased clustering and reduced diversity give \textbf{support to hypotheses 2.1a}. 

\begin{figure}
    \centering
    \begin{subfigure}[t]{0.50\textwidth}
        \centering
        \includegraphics[width=\textwidth]{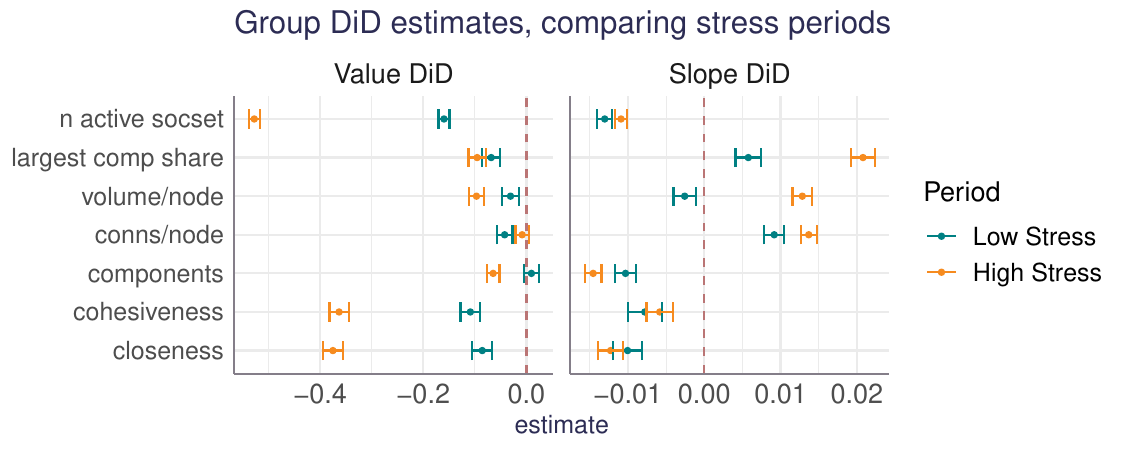}
        \label{fig:uncert_group}
    \end{subfigure}
    
    \hfill 
    
    \begin{subfigure}[t]{0.50\textwidth}
        \centering
        \includegraphics[width=\textwidth]{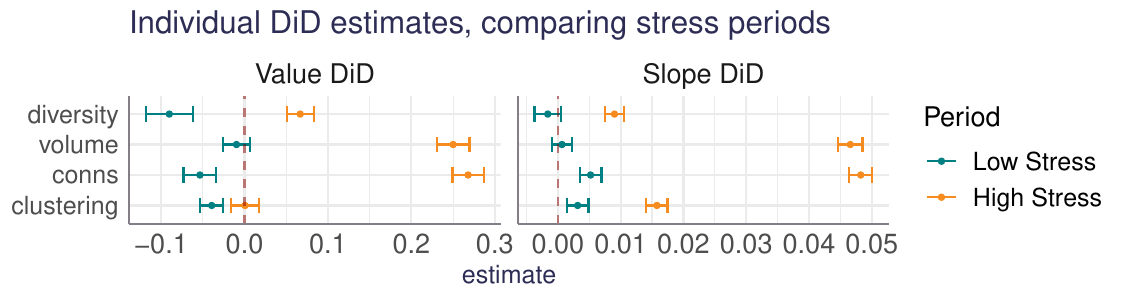}
        \label{fig:uncert_individual}
    \end{subfigure}
    
    \caption{Comparisons of estimates between periods of low and high stress. Top) Group perspective; larger effect sizes indicate increased communications breakdown under stress. Bottom) Individual perspective; under higher stress, individuals are more diverse and have more interactions}
\end{figure}

\subsection{\textbf{RQ2}, Response under heightened uncertainty} \label{sec:results:uncertainty}
In our investigation of RQ2, we delve into the responses of socialization sets for departing employees during two distinct periods. \cref{fig:uncert_group} displays overlapping value and slope DiD estimates for both periods, with a brighter orange for the period of increased stress. Our focus is to discern relative differences in these estimates.

Results in \cref{fig:uncert_group} suggest increased effect sizes in socialization patterns from the group perspective during periods of heightened uncertainty, evidenced by larger effect sizes in the metrics estimates that are roughly 3-4 times larger during high stress. \textbf{This supports hypothesis H2.1}. Group estimates typically align in direction (same sign) in the two periods, but distinct variations in magnitude emerge. Predominantly, group closeness and closure display significant negative value DiD estimates (-0.3756 DM and -0.3636 DM, respectively) under high stress. However, slope DiD estimates remain consistent across the periods. With the number of connected components, a noteworthy reversal appears, whereby the value DiD progresses from non-significant during a lesser stress period (0.0097 DM) to slightly negative under higher stress (-0.0646 DM3). For metrics of interaction intensity, the volume shows an increased drop during high stress (from -0.0309 DM to -0.0969 DM), but the group connections show comparatively an increase in magnitude from -0.0423 DM to non-significant during high stress. Pointing at groups attempting to maintain average connections during periods of higher stress, although at lower volumes.

The individual perspective during high stress displays interesting reversals in metrics \cref{fig:uncert_individual}, which support Hypotheses \ref{hyp:2.2a}. First, diversity presents a large reversal from negative (-0.0902 DM) to positive (0.0668 DM) under high stress. The individual clustering value DiD is negative pre-stress (-0.0393 DM), with a non-significant effect during high-stress. Individual connections and volume show minor negative estimates under less stress but a large positive value DiD for high stress. These estimates are around 4 times larger than the pre-stress estimates.

\subsection{\textbf{RQ3}, Heterogeneous effects} \label{sec:results:het}

\begin{figure}
    \includegraphics[width=0.45\textwidth]{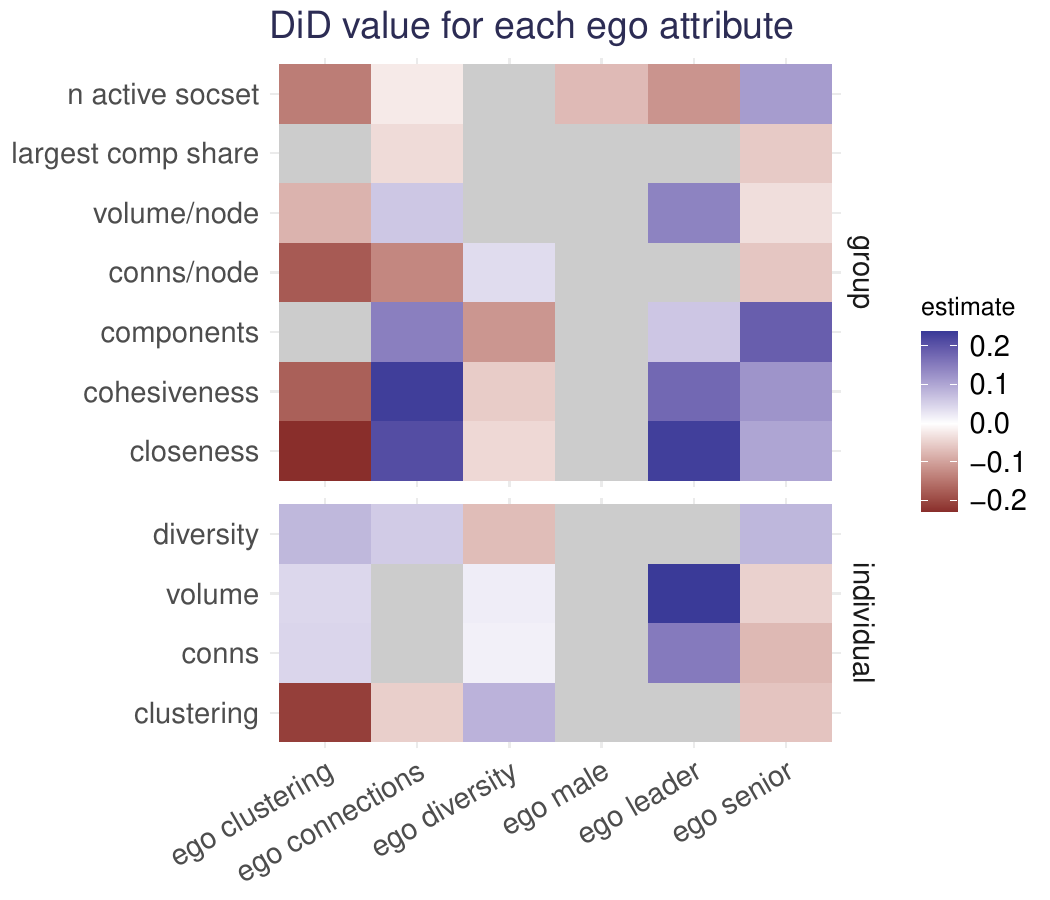}
    \Description[Value DiD for ego attributes]{Value DiD for ego attributes}
    \caption{Value DiD estimates from the ego attributes interaction model. Cell $(i, j)$ is the Value DiD comparing the levels of the ego attribute $j$ for metric $i$. For example, i=group cohesion, j=highly clustered ego vs. low-clustered has a value of  $~ -0.1$. Thus socialization sets where a highly clustered ego departs have decreased cohesion compared to a low clustered ego. Non-significant estimates in gray.}
    \label{fig:ego-attributes}
\end{figure}

In RQ3, we analyze nuances of impact within socialization groups, considering departing ego attributes. \cref{fig:ego-attributes} visually depicts DiD estimates from the ego attribute interaction model, with color intensity signifying the estimated value (darker hues represent higher values). For example, observe the cell corresponding to the DiD value for seniority contrast (column) and group cohesion metric (row). Here, DiD estimates enable the comparison of leaders against non-leaders (Details in \cref{app:modeling_details}). A positive coefficient, here ~0.1 DM, indicates that if the departing employee is a leader, the socialization set group cohesiveness is larger by 0.2 standard deviations of the metric (DM) than if the departing employee was a non-leader. In other words, compared to non-leaders, departing leaders leave socialization sets that end up more cohesive.

A few highlights: Higher departing ego clustering is associated with increased group communication breakdown, as evidenced by the negative values across group metrics. Higher ego diversity is associated with a further decline in communication metrics, but interestingly, the number of components decreases, meaning that the group becomes more connected but does not increase communication. Leader departing egos are associated with increased components but also increased group interaction intensity, pointing at group breakdown in communicative silos. Departing seniors are associated with increased components but with lower interaction intensity, pointing at less active silos. Regarding gender, our estimates find no support for the ego’s gender being associated with significant differential effects. That is, for this model, the change in slope and value is the same for socialization sets where the departing ego was male or female (data provided by the company only had a binary categorization of gender).

\section{Discussion}

\paragraph{Breakdown of socialization after employees departure}\label{sec:results_main}
We find a significant breakdown in the socialization set interactions after an employee's departure, supporting hypotheses \textbf{H1.1} and \textbf{H1.2a}. This disruption is characterized by a decrease in communication volume, connectivity, cohesion, and efficiency in the group perspective, leading to fragmentation into isolated silos. From the individual perspective, members tend to become less communicative, establish fewer connections, and exhibit increased clustering and lower diversity. These breakdown effects persist over time, despite a potential rebound to the status quo in some of the metrics.
Our results indicate that there is more to resignation than just losing the employee, given the loss of social capital enabled by their connections. This is particularly relevant in high collaboration contexts \cite{krepsCorporateCultureEconomic1990, elliottCorporateCultureOrganizational2023}. Our finding of reduced efficiency in communications of groups could explain the findings of other studies that observe a decline in performance in groups after resignation \cite{abbasiTurnoverRealBottom2000,sheehanEffectsTurnoverProductivity1993}. Our findings on the reduced interaction intensity in the individual perspective are also consistent with previous literature that finds departures affecting morale \cite{mujtabaLayoffsDownsizingImplications2020, brandEffectsLayoffsPlant2008} since interactions among colleagues also entail emotional support \cite{clarkegarciaTiesLeadersTeams2014a}. 
\paragraph{Potential mechanisms mediated by departing employee attributes}
Using the perspective of triadic closure \cite{granovetterStrengthWeakTies1973, podolnyResourcesRelationshipsSocial1997, clarkegarciaTiesLeadersTeams2014a}, we propose a potential mechanism behind the aforementioned breakdown. Suppose A, B, and C form a triangle. When A departs, the connection between B and C is weakened. Thus, it is more likely to observe B and C not communicating as frequently or at all in the future. Extrapolating to a socialization set, the departing employee holds several 'closures' between its neighbors. In our analysis of heterogenous effects, we observe that when departing egos are more clustered (more closures), there is a stronger communication breakdown of the group. Similarly, we also note that for highly diverse departing individuals that bridge different groups, the group displays increased connections and a reduction of silos but also lower cohesiveness and efficiency. In other words, the group attempts to reconnect the silos of information but has a less efficient structure. This suggests a mechanism of adaptation where the network attempts to recover its procedural connections and information flow \cite{burtCooperationNetwork2022, burtStructuralHolesSocial1992, bellingeriLinkNodeRemoval2020}.
\paragraph{Increased stress exacerbates group breakdown, but also individual advantage}\label{sec:dis_stress}
In studying socialization dynamics of departures during high-stress periods, we found that from the group perspective, the effect sizes related to communication breakdown after the departure of an employee are larger. However, interestingly, from the individual perspective, we find that individuals start to communicate with more connections and show higher structural diversity. This suggests that the departure of a connection leads to an individual's structural advantage. These findings are consistent with previous research where after departures, peers assume advantageous and information brokerage positions \cite{burtCooperationNetwork2022, podolnyResourcesRelationshipsSocial1997, sparroweSocialNetworksPerformance2001}.
\paragraph{Implications for organizations}
While concrete managerial suggestions might be difficult to propose as the details of employees matter, our research offers important takeaways into how organizations might maintain effective communication, especially during stressful times. Our study suggests benefits from considering the impact of departures not only on the immediate interactions lost but also the broader networks. This includes identifying groups at risk due to their close ties with departing employees who could leave behind communication gaps. Our findings indicate that while remaining employees may try to fill these gaps, it could lead to a less efficient communication structures. Thus, identifying and strengthening critical communication pathways in groups at risk of decoupling is essential.

\paragraph{Implications for broader network research}
Our research also informs, more generally, the impact of node removals in networks\cite{bellingeriLinkNodeRemoval2020}. Our results highlight the complex, dynamic responses to node removals by connecting the ideas of triads, cohesion, and the network's adaptation to maintain operations. They also reveal how sometimes outcomes are detrimental for the group but beneficial for individuals. This study can potentially enhance our comprehension of node removals across various types of networks.
\paragraph{Limitations}
The study has several limitations. First, we do not employ direct performance measures such as employee productivity. Instead, we rely on previous literature to describe potential consequences of particular network structures, such as associating increased individual structural diversity with increased advantage within the organization. Nonetheless, we note that understanding network changes provides insight into aspects that are usually harder to measure, such as the cohesion of a group. Second, we do not account for interactions of different types, such as formal and semiformal ties, or team boundaries. These categorizations have been applied in prior research to offer a richer view of network structure effects \cite{podolnyResourcesRelationshipsSocial1997, clarkegarciaTiesLeadersTeams2014a}. Although we recognize the benefit of this nuance, we were interested in a more general view of interactions as a first approach to our research questions. Finally, our approach does not establish causal relationships. We, however, use matching to give better contextualized estimates of the potential effects of departures.

\section{Conclusion and Future Work}

Our empirical study sheds light on the effects of a departure in the socialization of peers. This exploration rested upon a network perspective and utilized large-scale internal messaging data. Post-departure, we find evidence of a significant disruption in the socialization patterns of the remaining employees. The size of this reaction seems to be moderated by both external factors, such as periods of high stress, as well as ego-centric factors, such as the level of communication or seniority. Future studies could aim to establish the effect sizes on performance characteristics thought to be linked to social capital. There is also a clear opportunity to parse out the effect of resignations on different types of network ties according to content. Following this, a greater understanding could be sought to determine if the effects we observe can be generalized to other types of networks.

\begin{acks}
The authors gratefully acknowledge the support of the National Science Foundation awards \#2208662 and \#2313137.
\end{acks}

\bibliographystyle{ACM-Reference-Format}
\balance
\bibliography{sample-base}
\newpage

\appendix
\counterwithin{figure}{section}
\section{Methods Appendix}
\subsection{Networks construction}
\label{app:networks_construction}
\paragraph{Weights of interactions}
The communication between two employees $i,j \in \mathcal{V}$ is aggregated by week by summing the following weights:  any single instance of direct communication between two employees $i,j \in \mathcal{V}$ adds weight $w_{i,j}=1$. In addition, when employees $i$ and $j$ are part of any group interaction involving $k$ participants, $w_{i,j}=1/k$ is added. For example, in a 10-person meeting, a message adds weight $0.1$ to each pair of recipients regardless of who sent the message. Intuitively, in a group of $k$ people, a message is replicated $k$ times, but the importance is diluted across the $~ k(k-1)$ possible pairs. This gives a weight of roughly $k/(k(k-1)) \sim 1/k$ per pair.

As a robustness check, we applied a weighting technique where interactions $i \rightarrow j$ and $j \rightarrow i$ were made equal by taking the harmonic mean.  Thus, interactions in both directions needed to be present. The qualitative results as shown in \cref{app:results_appendix} are consistent with our main analysis concerning changes in the socialization set post-departure, suggesting that our findings are not merely a byproduct of one-sided interactions.

\subsection{Matching details}
\label{app:matching_details}

\begin{figure}
    \includegraphics[width=0.45\textwidth]{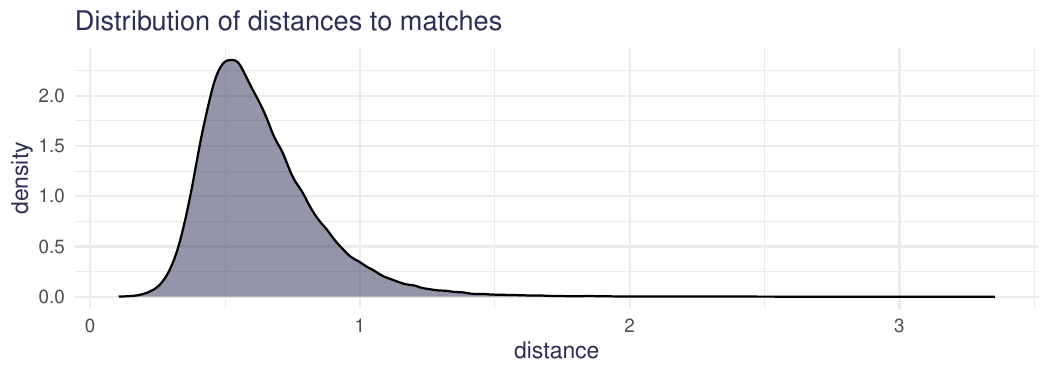}
    \Description[Distribution of weekly distances between treated and matches]{Distribution of weekly distances between treated and matches}
    \caption{Distribution of weekly distances between treated and matches. These distances are with respect to the space of the features used for matching.}
    \label{fig:match_qual_dist}
\end{figure}

For each employee $e$ that leaves the organization at a particular week, we select $m$ additional employees ${e'}_m$ to match $e$ with, based on similarity, from the set of employees who are not direct neighbors of $e$ in the communication set and were still present in the company at the time of $e$'s departure. We establish similarity both on the basis of the similarity of socialization set characteristics and similarity of the matched employee’s own communication characteristics. The reason for the latter is to account for behavioral variations of the egos that lead to different responses after resignation. For example, two departing employees may have an equal number of network neighbors but very different interaction intensities or different degrees of cohesion within their network. 

\paragraph{Matching features} Our similarity matching for the ego communication patterns uses the following set of attributes: The number of connections of the ego, volume of communications, clustering of the ego network, and a binary variable that denotes whether the employee is a manager or not. In other words, the individual perspective metrics of the ego that indexes the socialization set. Because computing the socialization sets and their metrics for all employees in the organization each week is computationally expensive, we typically used a proxy. The socialization set for an employee depends not only on their connections in that week, but rather a range of weeks.  Instead, we aggregate the connections by week, take the metrics on these (smaller) socialization sets, but then aggregate across weeks to obtain a more robust estimate. As we will note on the next section we still obtain qualitatively appropriate matches.

\paragraph{Matching procedure} To get a more stable estimate of the matching parameters for the departing ego and possible matches we aggregate the estimates over the weeks corresponding to the time interval of the freeze period used to identify socialization sets (ie. $[t^*-10, t^*-6]$). This leaves us with a dataset of averages of the aforementioned measures for each employee in the organization and for each week. This is the input for the matching procedure. We then use use $k$ nearest neighbors ($k$-NN) as to generate the top matches using the standardized features as inputs. For each departing employee $e$, we match $k$ non-departing employees on the same week that $e$ departed. 

A problem is that the same nodes $e'$ may be matched with many departing employees, creating artificial autocorrelation. This is especially problematic if the same node $e'$ is matched to several nodes departing around the same time.  Given that a significant portion of employees departed, such overlaps are likely by chance, and we take the following steps to alleviate this issue.   First we use $20$-NN to find the top 20 matches and then randomly select 3 matches from this set.  Second, we exclude from consideration nodes that were matched in the previous four weeks.

\begin{figure}
    \includegraphics[width=0.47\textwidth]{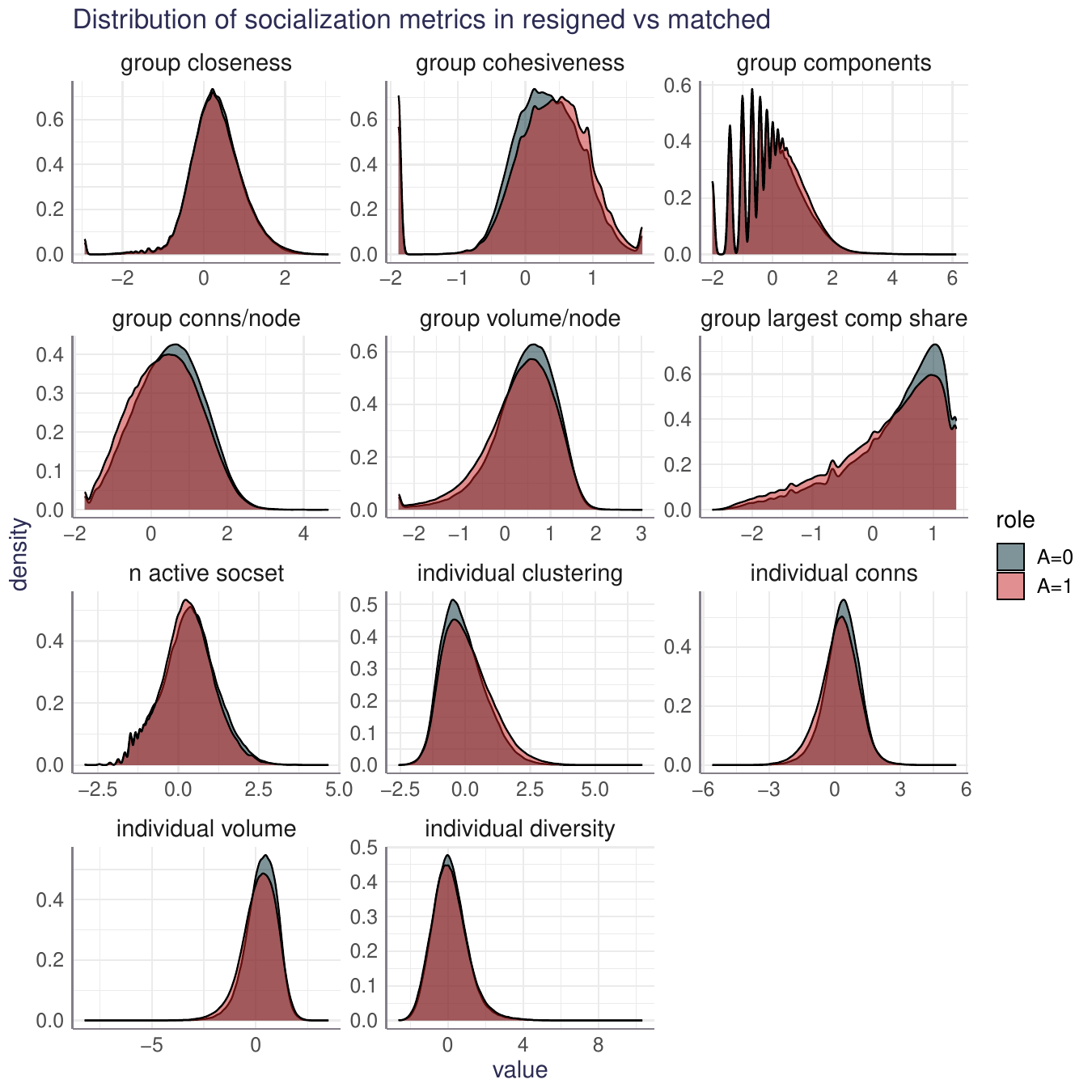}
    \Description[Distribution of weekly distances between treated and matches]{Distribution of weekly distances between treated and matches}
    \caption{Distribution of target metrics of the socialization set comparing treated and control socialization sets. The distribution plots only values before departure}
    \label{fig:match_qual_socsets}
\end{figure}

Then, with the matches selected, to assess match quality we inspect the distribution of distances to matches, \cref{fig:match_qual_dist}, and compare the distributions for the metrics of interest of the socialization sets among the resulted treated and control groups \cref{fig:match_qual_socsets}. We perform this comparison through visual inspection of the distributions.

\subsection{Modeling Details}
\label{app:modeling_details}

\paragraph{Mathematical definitions of Value DiD and Slope DiD}
We now provide mathematical definitions and computation details for the model estimates.

First, we define $DiD_{val}(\hat{f})$ by 
\begin{equation}
    DiD_{val}(\hat{f}) = D(\hat{f}^m|A=1) - D(\hat{f}^m|A=0)
\end{equation}

where $D$ corresponds to a difference between periods, within the same group and $A = 1$ indicates that this ego departed whereas $A = 0$ indicates is matched ego.

\begin{equation}
    D(\hat{f}^m| A=a) = \hat{f}^m(t=t^+| A=a) - \hat{f}^m(t=t^-| A=a).
\end{equation}

$D$ contrasts measurements from weeks $t^+ - t^-$, which are selected to be 8 weeks before and after departure within each of the groups. We select 8 weeks since pre-departure this falls in the middle of the period of the definition of the socialization set. 

Second, the estimate for the slope DiD

\begin{equation}
    DiD_{slp}(\hat{f}^m) = \partial_t D(\hat{f}^m|A=1) - \partial_t D(\hat{f}^m|A=0)
\end{equation}

captures, for each group metric, the difference between its slope pre and post intervention (departure). These estimates reveal relative metric changes and whether they increase or decrease over time. For instance, a negative $DiD_{val}$ and $DiD_{slp}$ for cohesion suggest that the decrease in cohesion is comparatively greater in the treated group, and additionally that this decrease worsens over time. We obtain these estimates with the R package \textit{emmeans} \cite{lenthEmmeansEstimatedMarginal2023}.

\paragraph{Target transformations} To facilitate a nuanced interpretation via comparison of change across the different metrics and model measures, we employ two transformations to our target metric variables. 

\begin{itemize}
    \item A log transformation $\log(f) + 1$ to address heavy-tailed behavior in the metrics volume, connections, and neighborhood size.
    \item z-scoring on our target variables to standardize the data, enabling us to compare the magnitude of changes across diverse metrics.
\end{itemize}
These quantities become the dependent variables $f^{m}$ in our models. Leveraging this approach allows us to contextualize metric alterations using a scale defined by standard deviations of the population. This enables the estimation of adimensional effect sizes and to compare effect sizes across different metrics.

\paragraph{Heterogeneous effects attributes}

Here we outline the definitions of the attributes that we used in the analysis for RQ3

\begin{description}
    \item $leader$ is a binary variable that signifies whether an ego assumes a leadership role. To determine this, we leverage data from the org chart during the months of July to August and assess whether the ego appears as a leader during this period. 
    \item $senior$ denotes whether a leader has more than five years of experience after college, corresponding to the 75th quantile of this variable in the data. This information is sourced from a company-provided dataset, which offers a snapshot of this data. 
    \item $gender$, in this dataset, is represented as a binary variable indicating whether an employee is registered as male or female within the company. Like seniority, this information is also derived from a single snapshot.
\end{description}

On the other hand, for communication attributes of the departing ego we use measures for the following: ego's volume of communication, number of connections, clustering and structural diversity. These measures are essentially the 'individual perspective' of the departing ego, as defined in \cref{sec:methods:measures}.

\paragraph{Heterogeneous effects model}
For analyzing heterogeneous effects, first, we only fit these models with data pertaining to socialization sets with indexing egos that have left the company. In other words, we restrict our analysis to the treated group. The model takes the following form 

\begin{equation}
f^{m}_{e,t} \sim X_e \times (t + hinge(t) + jump(t)) + controls_e +\eta_{e,t}
\end{equation}

Where $X_e$ denotes the set of attributes of the departing employee $e$. The model includes interactions between each ego attribute and the time basis.
For analysis, we generate marginalized difference-in-differences (diff-in-diff) estimates between levels of each ego attribute. This lets us contrast estimates between levels of the attributes. For instance, when examining gender, we compute the following estimate: 

\begin{equation}
    DiD(\hat{f}^m) = D(\hat{f}^m| gender=Male) - D(\hat{f}^m| gender=Female)
\end{equation}

where, 
 \begin{equation}
     D(\hat{f}^m| A=a) = \hat{f}^m(t=8| A=a) - \hat{f}^m(t=-8| A=a)
 \end{equation}

\paragraph{Level comparison estimates for heterogeneous effects}
For ego attributes that are included in the model as continuous variables, such as ego volume of communication, we generate estimates to compare the 1st and 3rd quartiles of the distribution. In essence, we investigate how lower-volume departing egos compare to higher-volume egos, as defined by the distribution. For the case of ego volume, the Diff-in-Diff estimate is computed as:

\begin{equation}
    DiD(f_m) = D(f_m| vol=Q3(vol)) - D(f_m| vol=Q1(vol))
\end{equation}

This analysis provides insights into the varying response contingent on different ego attributes on socialization network metrics during the defined time frame surrounding ego departures.

\section{Results Appendix}
\label{app:results_appendix}

\begin{figure}
    \includegraphics[width=0.485\textwidth]{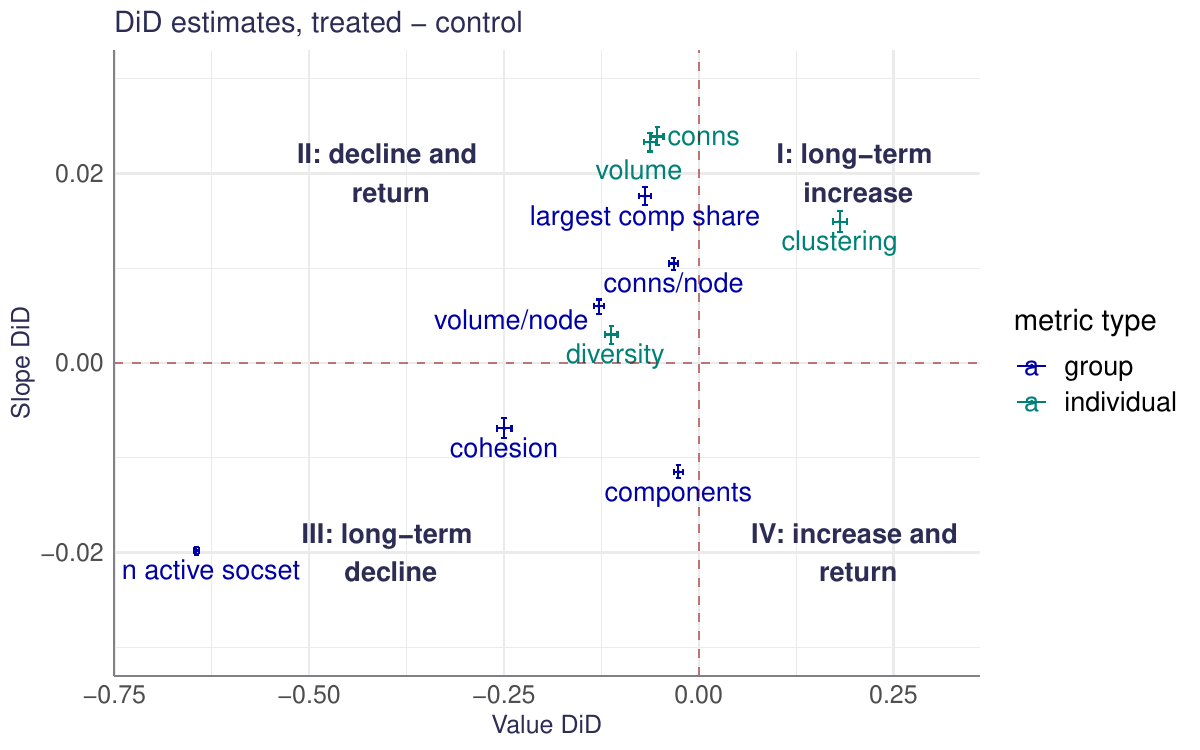}
    \Description[Robustness test for harmonic aggregation]{Robustness test for harmonic aggregation}
    \caption{Main result plot fr RQ1 using the harmonic weighting for aggregating directed edge weights. These results are qualitatively similar to \cref{fig:main}}
    \label{fig:robustness_wagg_harmonic}
\end{figure}

\begin{figure}
\begin{subfigure}{0.47\textwidth}
  \includegraphics[width=.97\linewidth]{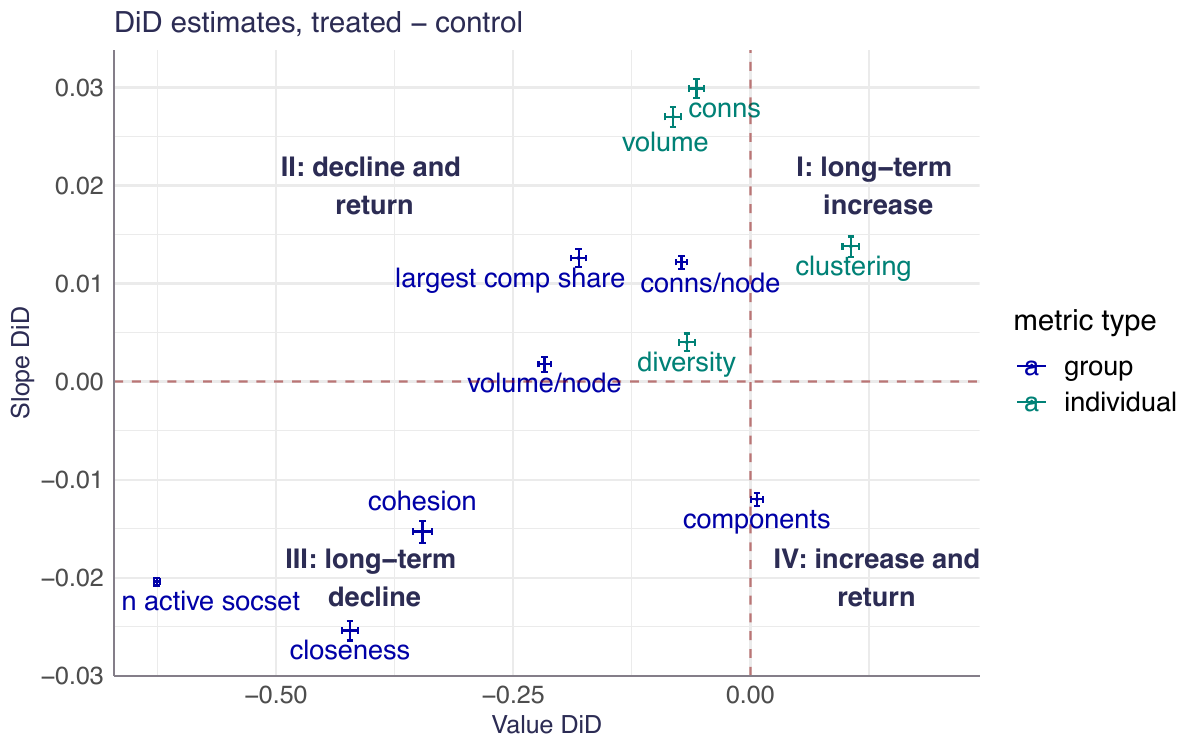}  
  \Description[Robustness test for n_freeze=5]{Robustness test for n_freeze=5}
  \caption{$n_{freeze} = 5$}
  \label{fig:robustness_freeze:5}
\end{subfigure}

\begin{subfigure}{0.47\textwidth}
  \includegraphics[width=.97\linewidth]{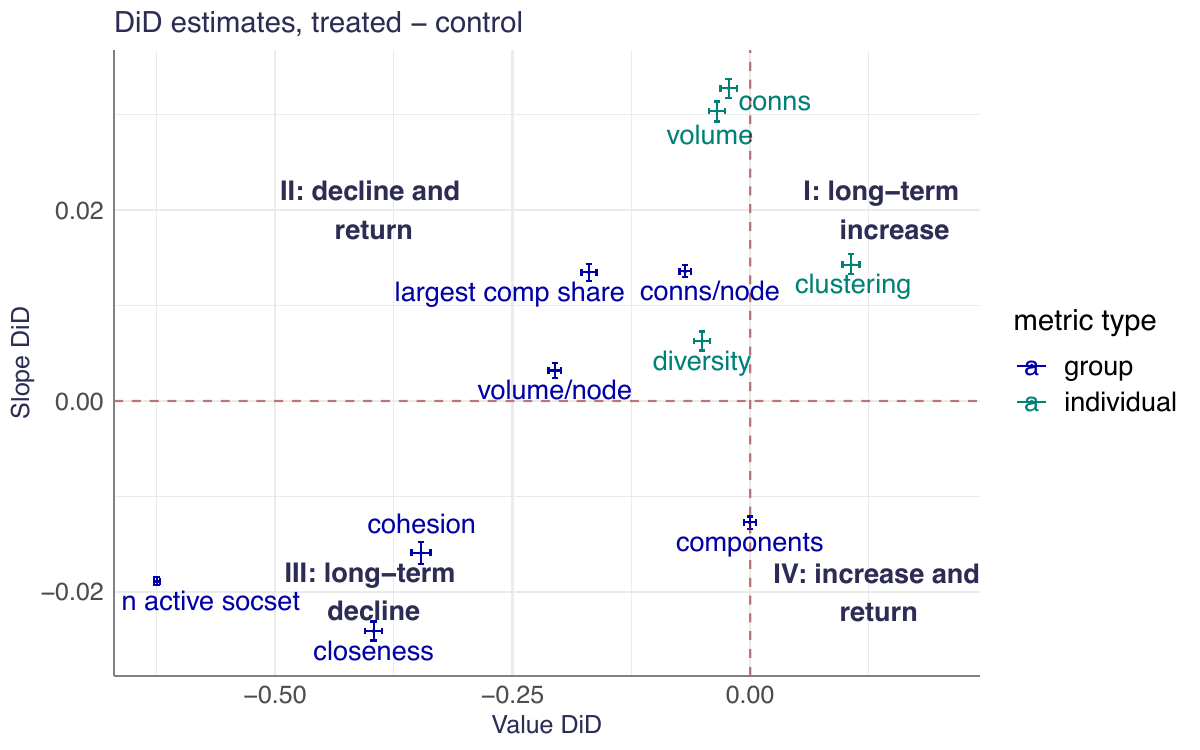}  
  \Description[Robustness test for n_freeze=6]{Robustness test for n_freeze=6}
  \caption{$n_{freeze} = 6$}
  \label{fig:robustness_freeze:6}
\end{subfigure}
\caption{Main result plot for RQ1 varying the freeze window that determines which contacts fall into the socialization set as described in \cref{sec:methods:networks}. These results are qualitatively similar to \cref{fig:main}}
\label{fig:robustness_freeze}
\end{figure}

\begin{figure}
    \includegraphics[width=0.485\textwidth]{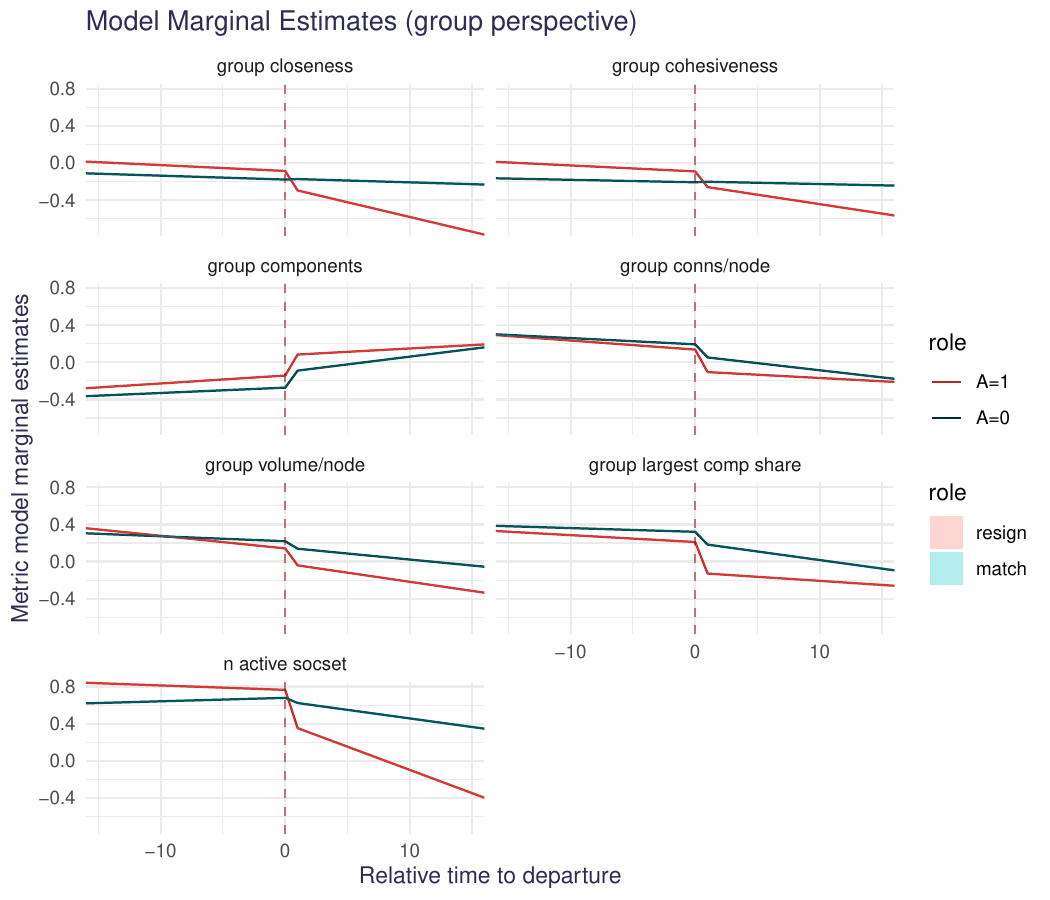}
    \Description[Marginal Estimates for RQ1]{Marginal Estimates for RQ1}
    \caption{Model marginal estimates of the metrics including the effect over time. Using the models for RQ1 as defined in \cref{sec:methods:model}.}
    \label{fig:emm-grp}
\end{figure}

\begin{figure}
    \includegraphics[width=0.485\textwidth]{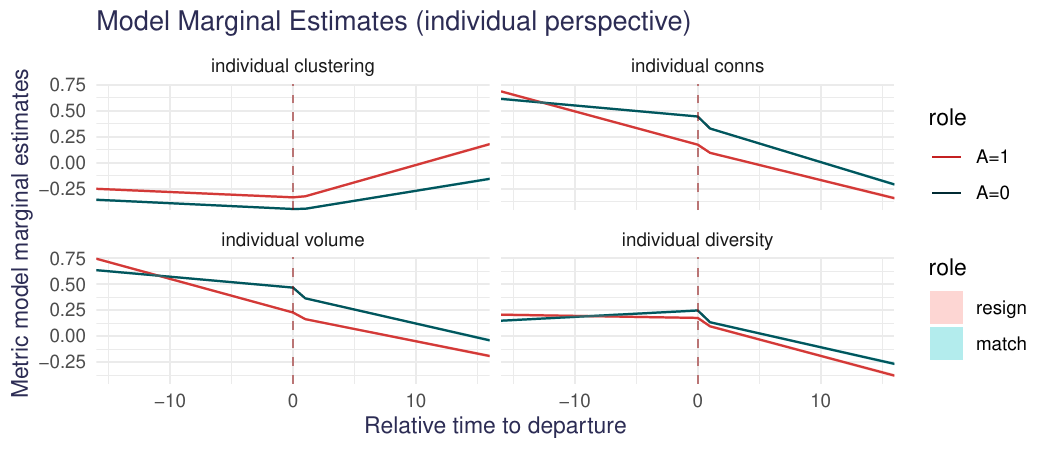}
    \Description[Marginal Estimates for RQ1]{Marginal Estimates for RQ1}
    \caption{Model marginal estimates of the metrics including the effect over time. Using the models for RQ1 as defined in \cref{sec:methods:model}.}
    \label{fig:emm-ind}
\end{figure}

Following, we list the plots concerning robustness tests. 
First, \cref{fig:robustness_wagg_harmonic} shows the robustness test varying the weighting scheme in the construction of the network. In this case, the weighting scheme is performed with the harmonic mean as described in \cref{app:networks_construction}. The estimates we find are qualitatively similar to using simple aggregation of weights. This means that the behavior we are observing in the networks is likely not a result of one-sided interactions. 
Second, \cref{fig:robustness_freeze} shows the main results plot varying the freeze window that is used in the definition of the socialization set, using longer values for the window of $n_{freeze}={5,6}$. The value used in the main paper uses a window size of $n_{freeze}=4$. Compared to the results in \cref{fig:main}, results for longer windows are qualitative similar. This means that the construction of the socialization sets is robust to variations in the weeks included, that include more possible contacts in the socialization sets.

As complementary results, we display marginal estimates of the metric models across time in \cref{fig:emm-grp} for the group perspective and \cref{fig:emm-ind} for the individual perspective. We provide these plots for reference of the time behavior of the metrics. However, our main analysis is conducted in the basis of the more simple Value DiD and Slope DiD estimates (defined in \cref{sec:methods:model}). As such, instead of comparing multiple estimates of the metrics, we abstract the comparison to relative changes in value and slope between the groups.

\end{document}